\documentclass[twoside,a4paper,11pt]{sea22}
\usepackage{graphicx}
\usepackage{hyperref}
\usepackage{media9}
\usepackage{amsmath}
\usepackage{subcaption}
\begin{document}
\pagenumbering{arabic}
\pagestyle{myheadings}
\thispagestyle{empty}
{\flushleft\includegraphics[width=\textwidth,bb=58 650 590 680]{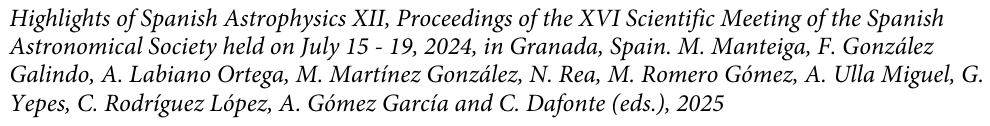}}
\vspace*{0.2cm}
\begin{flushleft}
{\bf {\LARGE
%
%%% TITLE of the paper. 
%%% TITLE of the paper. 
On the chlorine abundance in H~{\sc ii} regions
%
% Do not delete next few lines
}\\
\vspace*{1cm}
%
%%% Include here the LIST OF AUTHORS.
%%% Include here the LIST OF AUTHORS.
%%% Note that the last author has to be preceeded by an AND.
Orte-García, M.$^{1,2}$,
Esteban, C.$^{1,2}$,
Méndez-Delgado, J.E.$^{3}$,
and 
García-Rojas, J.$^{1,2}$
%
% Do not delete next few lines
}\\
\vspace*{0.5cm}
%
%%% AFFILIATIONS LIST.
%%% and the AFFILIATIONS LIST. Note that one affiliation per line.
%%% Add as many affiliations as necessary. 
$^{1}$
Instituto de Astrofísica de Canarias, E-38205 La Laguna, Tenerife, Spain\\
$^{2}$
Departamento de Astrofísica, Universidad de La Laguna, E-38206 La Laguna, Tenerife, Spain\\
$^{3}$
Astronomisches Rechen-Institut, Zentrum für Astronomie der Universität Heidelberg, Mönchhofstraße 12-14, D-69120 Heidelberg, Germany
%
% Do not delete next few lines
\end{flushleft}
%
% Headings
\markboth{
%%% Type the SHORT version of the paper title.
%%% Type the SHORT version of the paper title.
On the chlorine abundance in H~{\sc ii} regions
}{ % Do not delete
%
%%%  First Author \& Second Author   OR   First-author et al. 
%%%  First Author \& Second Author   OR   First-author et al. if the author list 
%%% contains three or more authors.
Orte-García, M. et al.
% 
% Do not delete next few lines
}
\thispagestyle{empty}
\vspace*{0.4cm}
\begin{minipage}[l]{0.09\textwidth}
\ 
\end{minipage}
\begin{minipage}[r]{0.9\textwidth}
\vspace{1cm}
\section*{Abstract}{\small
%
% ABSTRACT ABSTRACT ABSTRACT
% ABSTRACT ABSTRACT ABSTRACT
%%% Type the ABSTRACT of your oral contribution or poster

Chlorine (Cl) is a chemical element of the group of the halogens and is between the 17th and the 20th most abundant elements in the Solar System. It is thought to be produced from the capture of a proton or neutron by specific alpha-element isotopes during both hydrostatic and explosive oxygen burning, though some contribution may come from Type Ia supernovae. Cl lines are quite rare in stellar spectra, so most of the information available  about its abundance comes from analyzing the emission lines of ionized nebulae, especially the collisionally excited lines of Cl$^{2+}$ ([Cl~{\sc iii}] $\lambda\lambda$5518,5538). Our goal is to accurately determine the Cl abundance in H~{\sc ii} regions, and gather more information about its nucleosynthetic origin.
For this work we used a sample of observations that encompasses the deepest spectra of H~{\sc ii} regions available in the literature, from both the Milky Way and other galaxies in the local Universe, covering a range of oxygen (O) abundances, 12+log(O/H), from 7.18 to 8.70. As a first step, we determine the most representative electron temperature of the zone of the nebulae where the Cl$^{2+}$ ion lies. To this aim we used a grid of photoionization models and diagnostics valid for other ions, as that parameter cannot be determined directly through [Cl~{\sc iii}] lines. We then computed the total Cl abundance using different sets of ionization correction factors to account for the contribution from unseen ionization stages.
%
% Do not delete next few lines
\normalsize}
\end{minipage}
%
%
%%% BODY of the paper
%%% BODY of the paper
%
\section{Introduction \label{intro}}

H~{\sc ii} regions are nebulae photoionized by the ultraviolet radiation from young, massive O- and B-type stars, with effective temperatures ranging between 28,000 and 50,000 K. These nebulae are key to understanding the present-day chemical composition of the interstellar medium, shaped by stellar nucleosynthesis and Galactic chemical evolution. By studying their emission line spectra, we can derive both the physical conditions of the gas —such as electron density and temperature ($n_{\rm e}$, $T_{\rm e}$)— and the elemental abundances.

This study focuses on chlorine (Cl), a halogen group element, which is  approximately the 17th most abundant element in the Solar System \cite{asplund09}. Cl is mainly analyzed in H~{\sc ii} regions, because its spectral lines are extremely rare in stellar spectra, making nebular emission lines the main source of abundance data. The [Cl~{\sc iii}] lines at 5517 and 5537 $\text{\AA}$, the brightest Cl lines in the optical range, are typically used to determine the Cl abundance.

Under typical nebular conditions, Cl$^{2+}$ is the main contributor to the total Cl abundance, with smaller contributions of Cl$^{+}$ and Cl$^{3+}$. The lines of these last two ions, such as [Cl~{\sc ii}] $\lambda$9123 and [Cl~{\sc iv}] $\lambda$8046, are only detectable in the deepest spectra. In most cases, only [Cl~{\sc iii}] lines are available, so ionization correction factors (ICFs) are needed to account for the unseen ions. Additionally, the $T_{\rm e}$ of the Cl$^{2+}$ zone cannot be easily determined, so we rely on temperature relations and diagnostics for other ions to estimate it. This is the main aim of our study.

\section{Photoionization models and observational sample}

To achieve the objectives of this project, we used photoionization models to simulate the spectra of H~{\sc ii} regions and compute their physical properties. Specifically, we employed the BOND photoionization model grid \cite{valeasari16}, available in the \textit{Mexican Million Models database (3MdB)} \cite{morisset15}, and applied a series of constraints to select the most appropriate models for our observational sample. The analysis of the model outputs was conducted using the {\sc PyNeb} package  \cite{luridiana}, which calculates $n_{\rm e}$, $T_{\rm e}$, and ionic abundances from the modeled spectra.

In addition to our primary goal, we aim to verify the applicability of certain temperature relationships to observational data. For this, a sample of 148 H~{\sc ii} regions and 120 star-forming galaxies with detected [Cl~{\sc iii}] lines has been used. This sample constitutes the largest sample of ionized nebulae with reliable determinations of Cl abundances. We include our own observations (see compilation in \cite{mendezdelgado2023}) and additional high-quality spectra from the literature. For each region, we reported line intensities and recomputed $n_{\rm e}$, $T_{\rm e}$ and ionic and elemental abundances.

To determine the Cl$^{2+}$ abundance, we initially used $T_{\rm e}$(S$^{2+}$), due to their similar ionization potentials. However, when no [S~{\sc iii}] diagnostic lines are available, we used a temperature relationship \cite{garnett1992}. This brought us back to the models and our main goal: identifying a reliable temperature relationship for Cl$^{2+}$. Instead of relying on S$^{2+}$, we explored a relationship between $T_{\rm e}$(Cl$^{2+}$) and $T_{\rm e}$(O$^{2+}$) and $T_{\rm e}$(N$^{+}$) obtained from the models, chosen because the ionization potential of Cl$^{2+}$ is intermediate between to those of O$^{2+}$ and N$^{+}$, and $T_{\rm e}$(O$^{2+}$) is typically measured with high accuracy through the classical [O~{\sc iii}] $\lambda\lambda$4959,5007$\lambda$4363 line ratio in most H~{\sc ii} regions.

\section{Temperature relationships}

\begin{figure}[htb]
    \centering
    \begin{subfigure}[b]{0.49\textwidth}
        \centering
        \includegraphics[width=\textwidth]{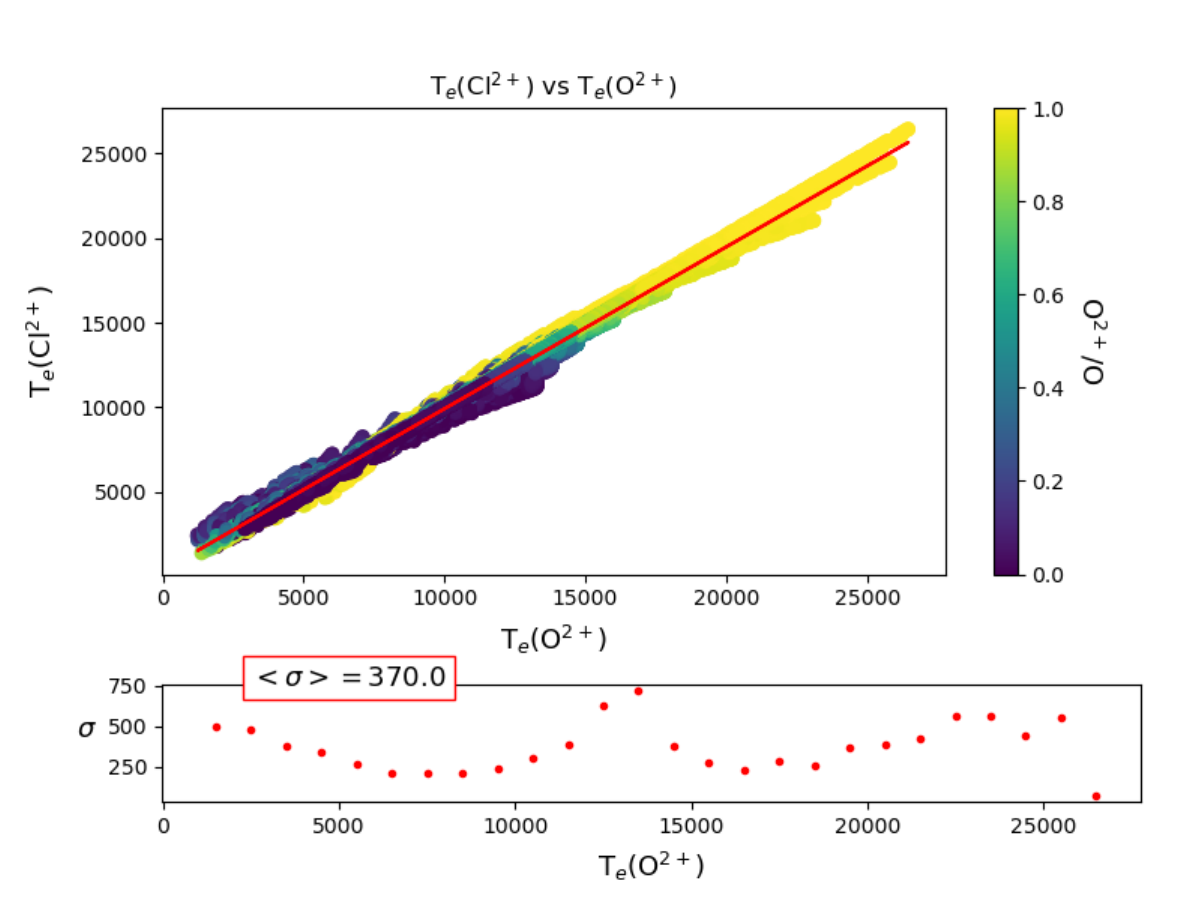}
        \caption{$T_{\rm e}$(O$^{2+}$)}
        \label{fig1}
    \end{subfigure}
    \hfill
    \begin{subfigure}[b]{0.49\textwidth}
        \centering
        \includegraphics[width=\textwidth]{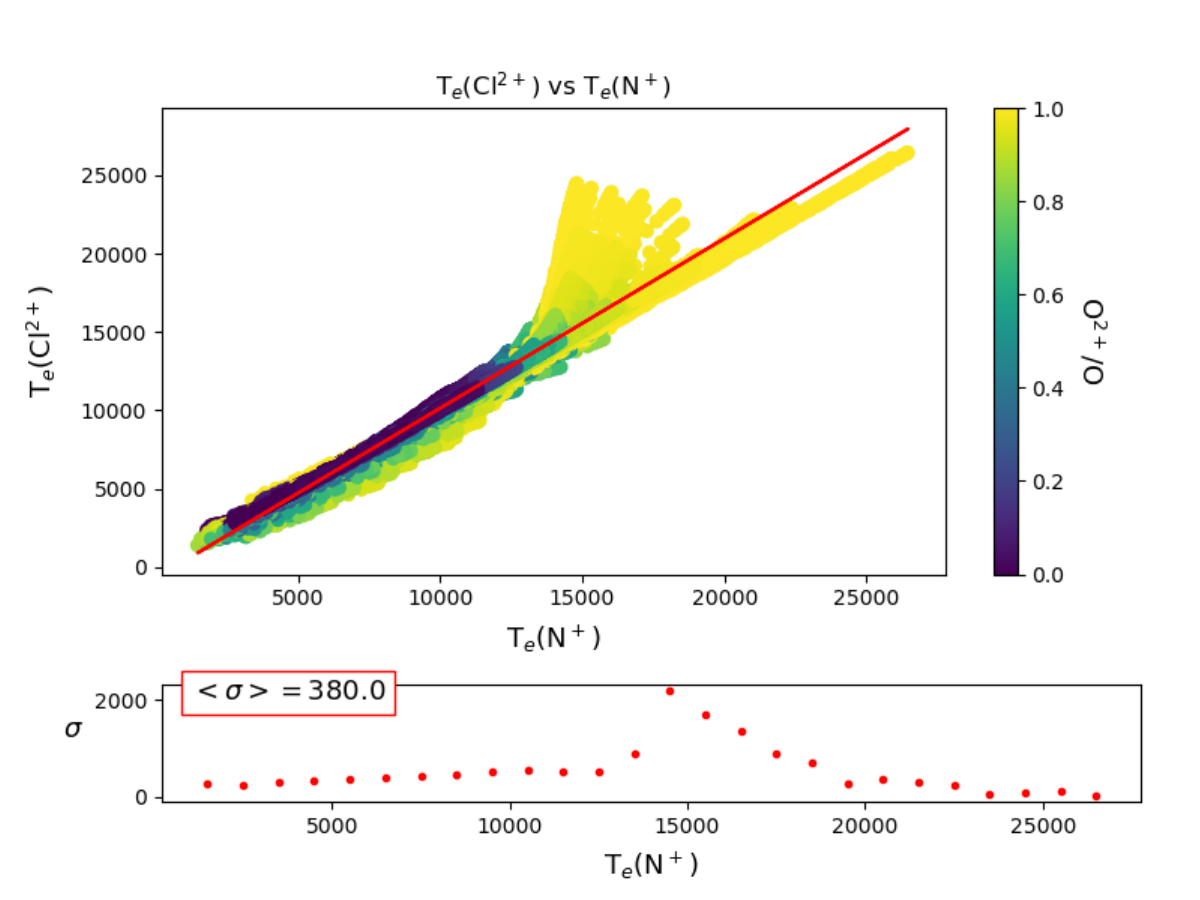}
        \caption{$T_{\rm e}$(N$^{+}$)}
        \label{fig2}
    \end{subfigure}
    \caption{$T_{\rm e}$(Cl$^{2+}$) vs. (a) $T_{\rm e}$(O$^{2+}$) and (b) $T_{\rm e}$(N$^{+}$) obtained from photoionization models. The points are color-coded by its ionization degree. The red line is a linear fit to the points. In the bottom panel, we show the values of the standard deviation, $\sigma$, to the fit in bins of 1\,000 K. The mean value of $\sigma$ is shown inside the red box.}
    \label{fig12}
\end{figure}

To derive the temperature relationship we are looking for, we used the photoionization model grid to plot $T_{\rm e}$(Cl$^{2+}$) against $T_{\rm e}$(O$^{2+}$) and $T_{\rm e}$(N$^{+}$), as shown in Figs.~\ref{fig1} and ~\ref{fig2}, respectively. The observed behavior is: N$^{+}$, which represents regions with lower ionization, fits better at lower temperatures, while O$^{2+}$ provides a better fit in higher ionized regions.

The key conclusion from this analysis is that neither temperature seems to be appropriate for the whole range of ionization degrees, leading us to propose a  representative temperature ($T_{rep}$), defined as:
\begin{equation}
\label{trep}
    T_{rep}=T_{\rm e}(\rm N^+)\cdot\left(1-\frac{O^{2+}}{O}\right)+T_{\rm e}(\rm O^{2+})\cdot \frac{O^{2+}}{O}
\end{equation}

\begin{figure}[!h]
\center
\includegraphics[width=0.7\textwidth]{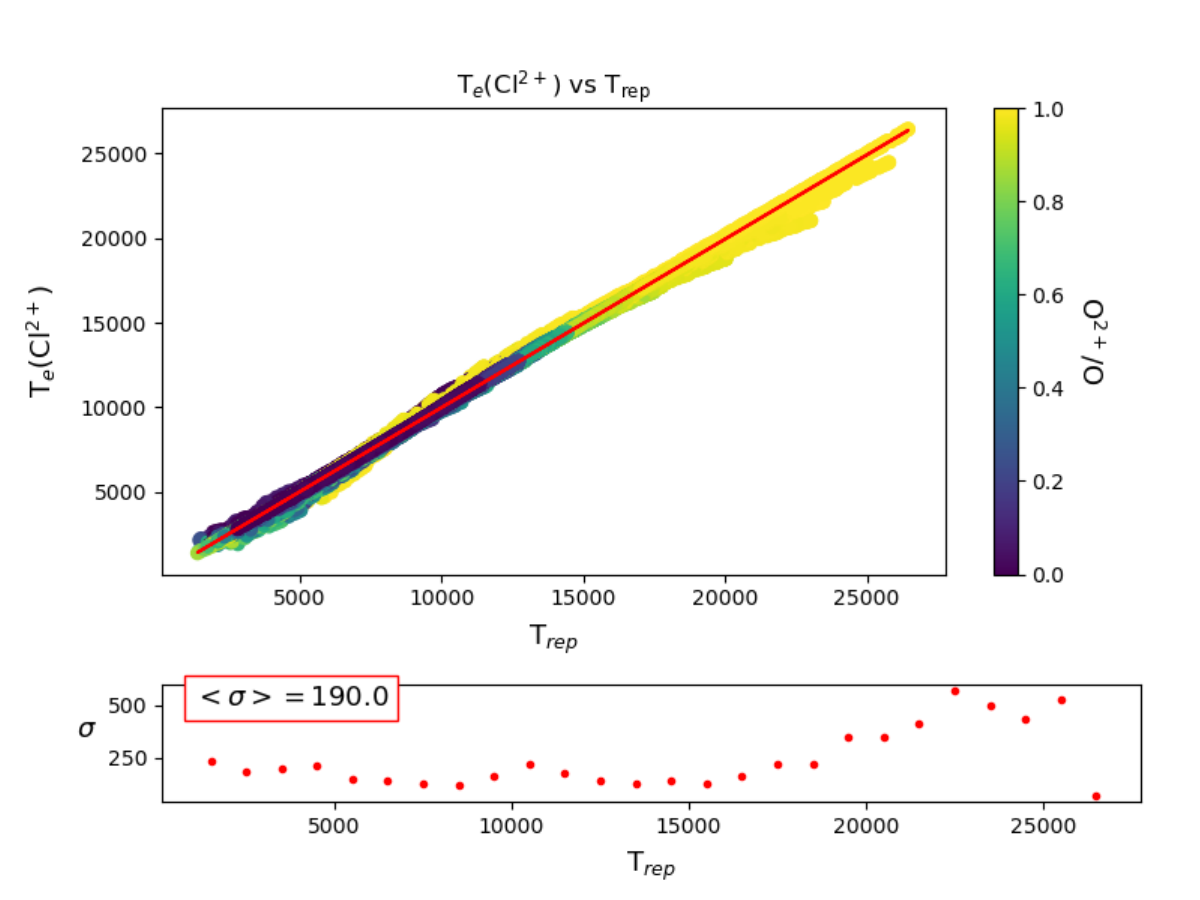}
\caption{\label{fig3} $T_{\rm e}$(Cl$^{2+}$) vs. $T_{rep}$(O$^{2+}$), from photoionization models. Color-code, lines and panels are equivalent to those shown in Fig.~\ref{fig12}.}
\end{figure}

This captures the trend observed in Fig.~\ref{fig12}: for lower ionization degrees, $T_{\rm e}$(N$^{+}$) has more weight, while  $T_{\rm e}$(O$^{2+}$) dominates at higher ionization. Re-plotting with this new temperature yields Figure~\ref{fig3}, which shows an improved behavior across ionization extremes and a significantly reduced standard deviation. Henceforth, we will compare results from our observational sample using this representative temperature.

\section{Cl abundances}

Once we have calculated the appropriate physical conditions, the next step is to complete the chemical abundance analysis of our observational sample using {\sc PyNeb} \cite{luridiana}. We started with the determination of the Cl$^{2+}$ ionic abundance. We then calculated the total Cl abundance using two different ICF schemes \cite{amayo2021, izotov2006}, one of which directly calculates the Cl/O ratio. This ratio is of interest because, Cl and O are both thought to be formed in hydrostatic and explosive oxygen burning in massive stars. Moreover, previous studies suggest that this ratio should remain constant and close to the solar value \cite{lodders2023}.

\begin{figure}[htb]
    \centering
    \begin{subfigure}[b]{0.49\textwidth}
        \centering
        \includegraphics[width=\textwidth]{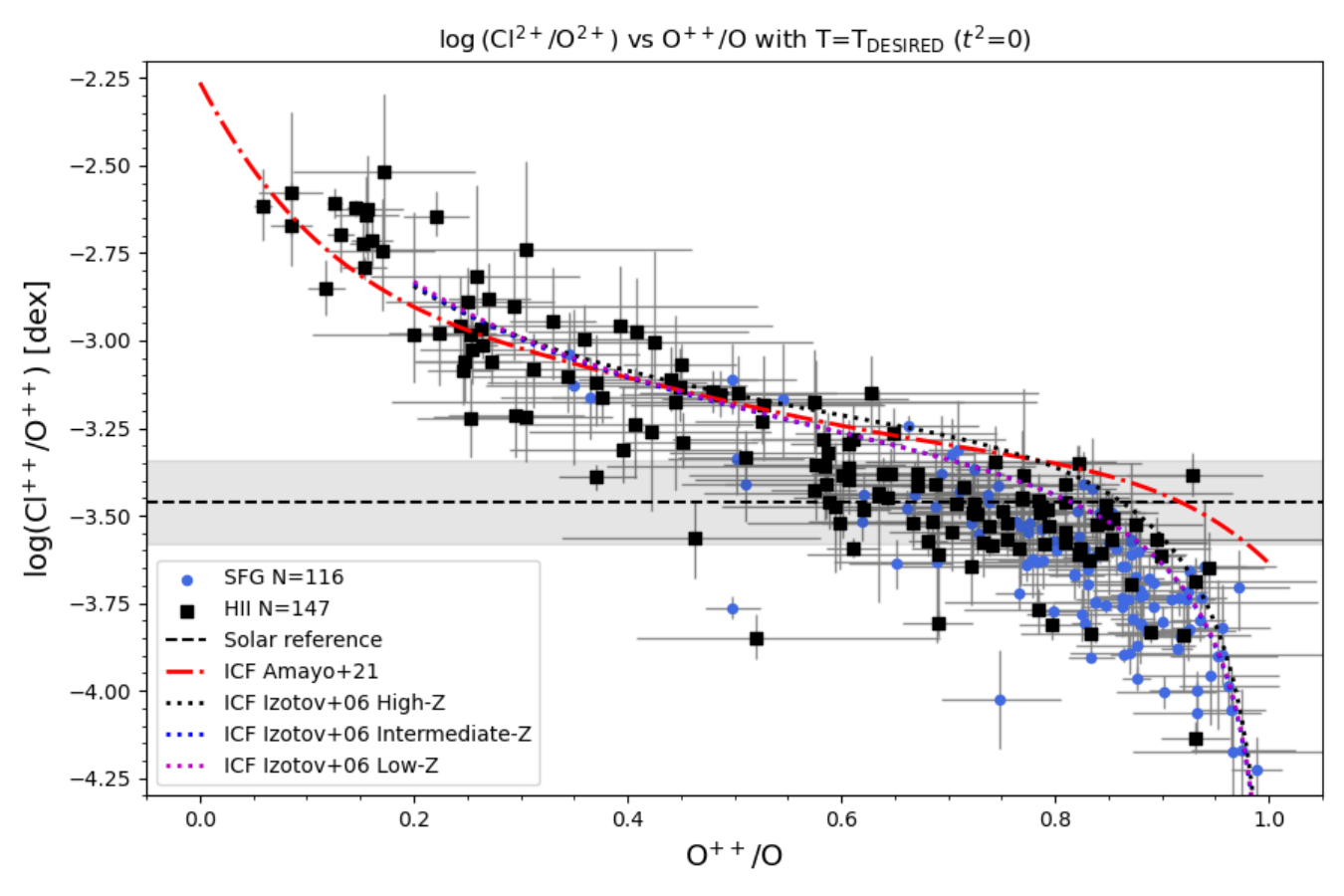}
        \caption{$T_{\rm e}$(S$^{2+}$)}
        \label{fig4}
    \end{subfigure}
    \hfill
    \begin{subfigure}[b]{0.49\textwidth}
        \centering
        \includegraphics[width=\textwidth]{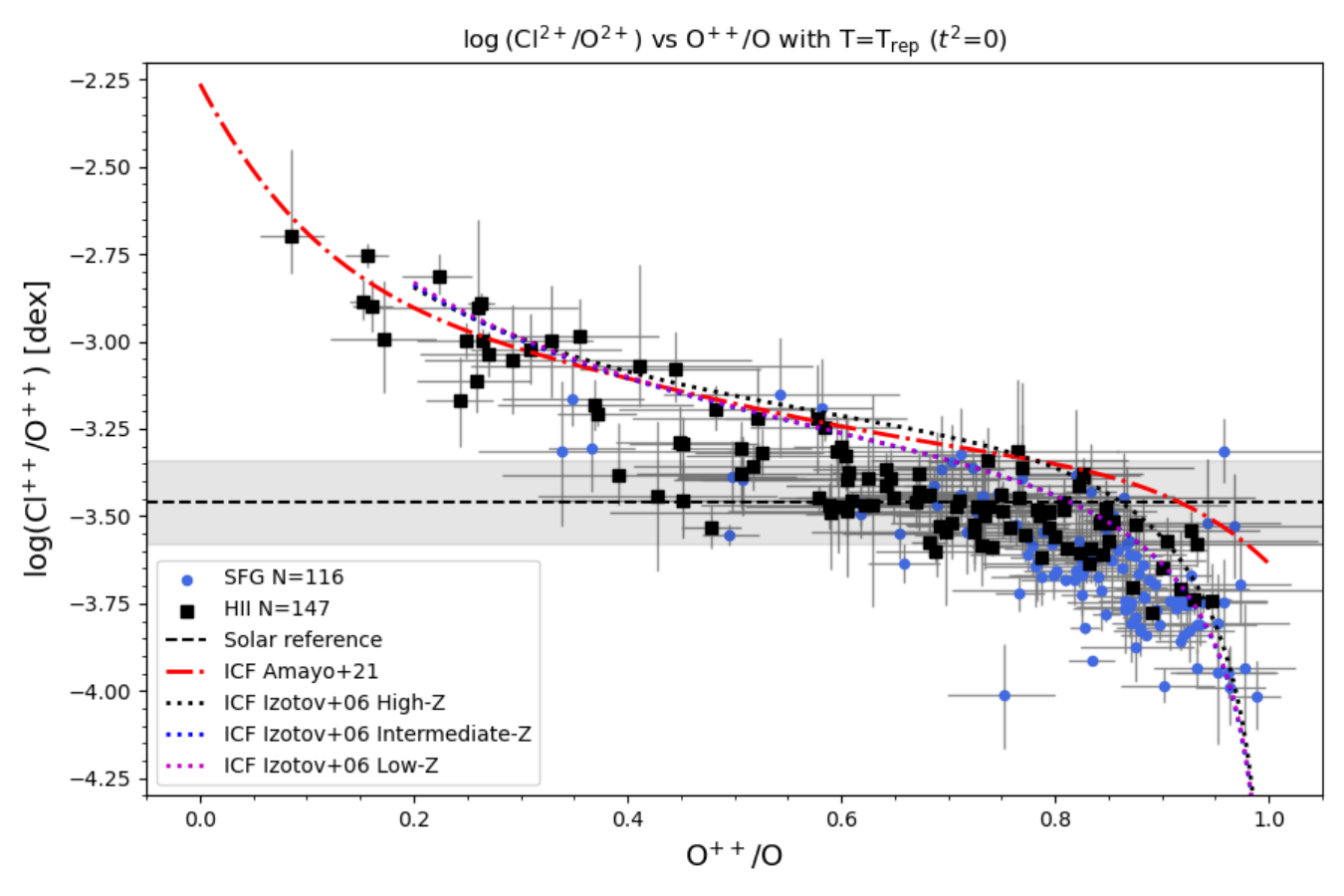}
        \caption{$T_{\text{rep}}$}
        \label{fig5}
    \end{subfigure}
    \caption{Cl$^{2+}$/O$^{2+}$ as a function of the ionization degree, calculated using: (a) $T_{\rm e}$(S$^{2+}$), and (b) $T_{\text{rep}}$ for the observational sample. In both figures, the blue points correspond to star forming galaxies, while the black ones are H~{\sc ii} regions. The dashed horizontal line marks the Solar Cl/O used as a reference, and the dotted lines display the curves of the different ICFs tested.}
    \label{fig45}
\end{figure}

We first plotted the Cl$^{2+}$/O$^{2+}$ ratio as a function of the ionization degree for our sample, alongside the functions of the different ICFs, for both $T_{\rm e}$(S$^{2+}$) and $T_{\text{rep}}$ (see Figs.~\ref{fig4} and~\ref{fig5}, respectively). The results show that the ICF from \cite{amayo2021} is valid across the full ionization range, while the ICF proposed by \cite{izotov2006} becomes applicable only when the ionization degree exceeds 0.2. However, at higher ionization degrees, the ICF of \cite{izotov2006} fits the observational trend better, while the ICF from \cite{amayo2021} does not.

\begin{figure}[!h]
\center
\includegraphics[width=\textwidth]{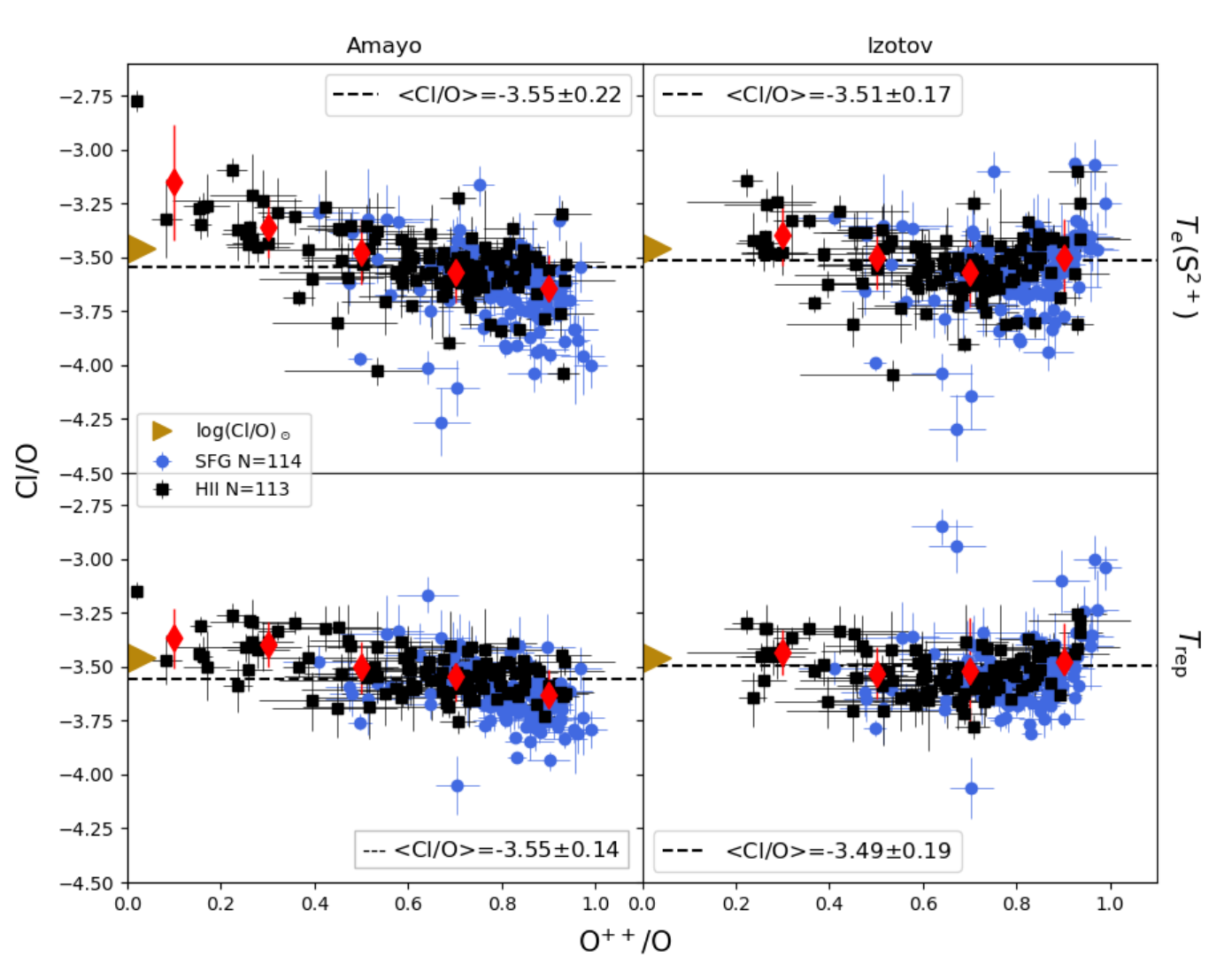}
\caption{\label{fig6} Cl/O as a function of the ionization degree for the sample objects. The left and right columns correspond to the ICF by \cite{amayo2021} and \cite{izotov2006}, respectively. The top and bottom rows show the results for $T_{\rm e}$(S$^{2+}$) and $T_{\text{rep}}$, respectively. Blue points correspond to star forming galaxies and black ones to H~{\sc ii} regions. The dashed horizontal line marks the mean Cl/O of the points, the red diamonds show the mean values for 0.2 wide bins, and the golden triangle indicates the Solar Cl/O used as a reference.}
\end{figure}

Finally, using the ICF equations, we calculated and plotted the Cl/O ratio as a function of the ionization degree, resulting in the plots shown in Fig.~\ref{fig6}. We observe that $T_{\text{rep}}$ tends to reduce the data dispersion, and that the ICF from \cite{izotov2006} gives smaller amplitudes in the Cl/O variations as a function of ionization degree, except at the highest ionization degrees. The peak on the right-hand side, more pronounced with $T_{\text{rep}}$, could be due to the relationship between the temperature and ionization degree. Using $T_{\rm e}$(S$^{2+}$) introduces this dependence when calculating the total Cl abundance, while $T_{\text{rep}}$ introduces it at the Cl$^{2+}$ abundance calculation stage, which might increase the dispersion in these points and, thus, the overall dispersion of the mean value. The key takeaway is that none of the combinations fully eliminates the dependence on either temperature or ionization degree.

%When representing the Cl/O ratio as a function of O/H abundance, we obtained new graphs showing the same points, along with the mean value (dashed line) and the solar reference value. We also included 33 stellar spectra points, represented by green stars. In this case, the dependence of the Cl/O ratio on O abundance allows us to explore nucleosynthesis, but we find that it also depends on the chosen methodology. Furthermore, these results reaffirm the flattening effect when using T$_{\text{rep}}$.

%Lastly, we performed a quick analysis of the Cl/S ratio, using the ICF of Izotov and T$_{\text{rep}}$. The resulting figure, plotting the ratio as a function of ionization degree, shows much higher dispersion, likely due to the stronger dependence on ionization degree introduced by the temperature criterion. Again, we observe the peak at the high end of the ionization range, even though the Izotov ICF performed better in the earlier analysis for high ionization degrees. Despite the increased dispersion, the mean value of the points remains close to the reference value.

\section{Preliminary conclusions and future work}

The preliminary conclusions of this study are, first, that the proposed representative temperature ($T_{\text{rep}}$) is valid for representing the region where Cl$^{2+}$ is located. Second, despite our expectations, none of the combinations of ICF and temperature used in this analysis completely eliminates the dependence on either temperature or ionization degree. Lastly, and most importantly, as we will describe in detail in future publications, the results seem to rule out significant contributions apart from the hydrostatic and explosive oxygen burning in massive stars in the nucleosynthesis of chlorine.

As for next steps, we plan to further investigate the ionization degree dependence introduced by the chosen temperature. Additionally, we are exploring alternative temperature relationships applicable to other ions, which is a key focus of our ongoing work.

%
%
% Do not delete the next line
\small  % Do not delete
%
%%% Comment the following line if you do not have acknowledgments.
\section*{Acknowledgments}   % Do not delete if you declare acknowledgments
%
%%% ACKNOWLEDGMENTS
%%% ACKNOWLEDGMENTS

%Since this work is a result of my Master's Degree Thesis, I would like to thank my tutors, César Esteban and José Eduardo Méndez Delgado, for their help and guidance during the whole research. I would also like to mention my actual PhD supervisor, Jorge García Rojas, who, along with César, has guided me in the presented next steps of the work, and my first months as a PhD student.

We acknowledge financial support from the Agencia Estatal de Investigaci\'on of the Ministerio de Ciencia e Innovaci\'on (AEI- MCINN) under grant ``Espectroscop\'ia de campo integral de regiones H II locales. Modelos para el estudio de regiones H II extragal\'acticas'' with reference DOI:10.13039/501100011033.
%
% Do not delete the next few lines

%
\end{document}